\let\Xdocument\document
\documentclass{iau}
\let\document\Xdocument

\usepackage{amsmath}
\usepackage{graphicx}
\usepackage{multirow}

\begin{document}

\lefttitle{S. F. Dermott, D. Li, \& A. A. Christou}
\righttitle{Deconvolving the complex structure of the asteroid belt}

\jnlPage{1}{7}
\jnlDoiYr{2021}
\doival{10.1017/xxxxx}

\aopheadtitle{Proceedings Kavli-IAU Symposium No. 382, 2024}
\editors{A. Lemaitre \& A. S. Libert, eds.}

\title{Deconvolving the complex structure of the asteroid belt}

\author{Stanley F. Dermott$^1$, Dan Li$^2$, and Apostolos A. Christou$^3$}
\affiliation{$^1$Department of Astronomy, University of Florida, Gainesville, FL 32611, US
\\ email: \email{sdermott@ufl.edu} 
\\ $^2$NSF’s National Optical-Infrared Astronomy Research Laboratory, Tucson, AZ 85719, US 
\\ email: \email{dan.li@noirlab.edu} 
\\ $^3$Armagh Observatory and Planetarium, College Hill, Armagh, BT61 9DG
\\ email: \email{Apostolos.Christou@armagh.ac.uk}}

\begin{abstract}
The asteroid belt is a unique source of information on some of the most important questions facing solar system science. These questions include the sizes, numbers, types and orbital distributions of the planetesimals that formed the planets, and the identification of those asteroids that are the sources of meteorites and near-Earth asteroids.  Answering these questions requires an understanding of the dynamical evolution of the asteroid belt, but this evolution is governed by a complex interplay of mechanisms that include catastrophic disruption, orbital evolution driven by Yarkovsky radiation forces, and chaotic orbital evolution driven by gravitational forces. While the timescales of these loss mechanisms have been calculated using estimates of some critical parameters that include the thermal properties, strengths and mean densities of the asteroids, we argue here that the uncertainties in these parameters are so large that deconvolution of the structure of the asteroid belt must be guided primarily by observational constraints. We argue that observations of the inner asteroid belt indicate that the size-frequency distribution is not close to the equilibrium distribution postulated by \citet{Dohnanyi1969}. We also discuss the correlations observed between the sizes and the orbital elements of the asteroids. While some of these correlations are significant and informative, others are spurious and may arise from the limitations of the Hierarchical Clustering Method that is currently used to define family membership.
\end{abstract}


\maketitle

\section{Introduction}

We discuss three classes of family asteroids. (1) Family asteroids that were formed either by catastrophic disruption or crater formation. (2) Halo asteroids that are members of known families. These are family members that, due to the limitations of the Hierarchical Clustering Method (HCM), have not been attached, unambiguously, to a specific family. (3) Asteroids in ghost families. These are older families that were formed in the same way as those in class (1) but are difficult to detect because their orbital elements have dispersed. If we removed all the asteroids in these three classes, then we would be left with the primordial asteroids. We note, however, that each family must have, or have had, a precursor asteroid that is the root source of the other family members. These precursor asteroids are also members of the primordial class. Answering many of the questions facing solar system science requires a determination of the structure of the primordial asteroid belt, that is, the structure of the belt after the transport of the asteroids from their formation locations to the main belt, and after the migration of the major planets that triggered that transport had ceased, but before the time of formation of any of the families. Determining that primordial structure and its transformation into the current structure are the challenges discussed in this paper.

We confine our discussion to the inner main belt (IMB) defined here by 2.1 $au$ $< a <$ 2.5 $au$ and $I < 18$ $deg$, where $a$ is the proper semimajor axis, and $I$ is the proper inclination. Family membership is usually defined using the Hierarchical Clustering Method (HCM) introduced by \citet{Zappala.et.al1990}. This uses the metric
\begin{equation}
	d= na\sqrt{d_1^2+d_2^2+d_3^2}
	\label{eq:1}
\end{equation}
where
\begin{equation}
	na=\sqrt{GM_{\odot}/a},\, d_1= \sqrt{5/4}(\Delta a/a),\, d_2= \sqrt{2}\Delta e,\, d_3= \sqrt{2}\Delta(\textrm{sin}I)
	\label{eq:2}
\end{equation}
and $na$ is the average orbital velocity of an asteroid having semi-major axis $a=(a_1+a_2)/2$, $\Delta a=a_1-a_2$, $\Delta e=e_1-e_2$, $\Delta \textrm{sin}I=\textrm{sin}I_1-\textrm{sin}I_2$, where the indices 1 and 2 denote the two bodies whose mutual distance, $d$ is calculated. Each member of a given family is separated from its nearest neighbor by $d<d_{\textrm{crit}}$, where $d_{\textrm{crit}}$ is a critical distance chosen such that the members of one family are not linked to the members of another family. Asteroids in the IMB with absolute magnitude $H < 16.5$ are observationally complete \citep{Dermott.et.al2018, hendler.et.al2020}. However,  while this set of asteroids is devoid of observational selection effects, some correlations between the sizes and the orbital elements of the family asteroids may arise because of  the way in which the HCM defines family membership.

\section{Orbital element and size correlations in the IMB}

\begin{figure}
	\centering
    	\includegraphics[width=1.74in]{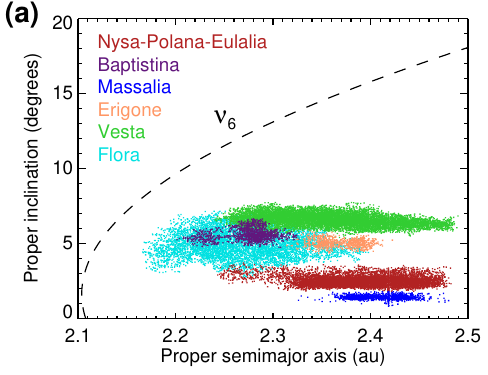}
	\includegraphics[width=1.74in]{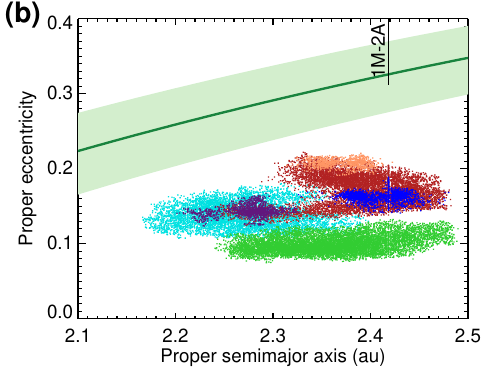}
	\includegraphics[width=1.74in]{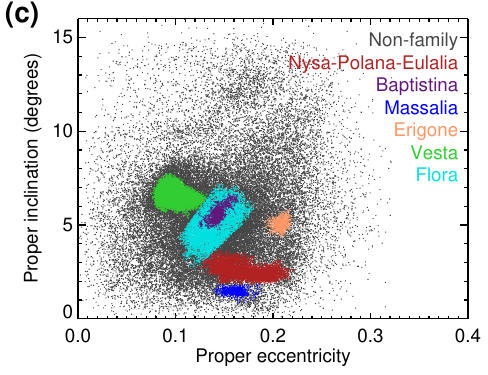}
	\includegraphics[width=1.74in]{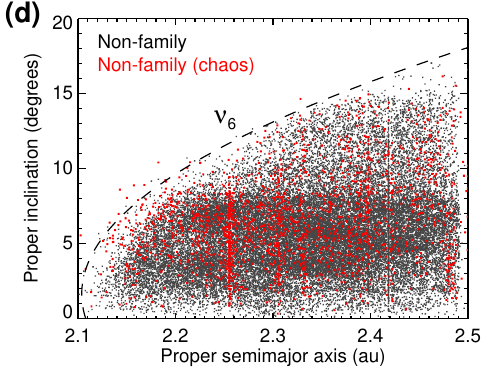}
	\includegraphics[width=1.74in]{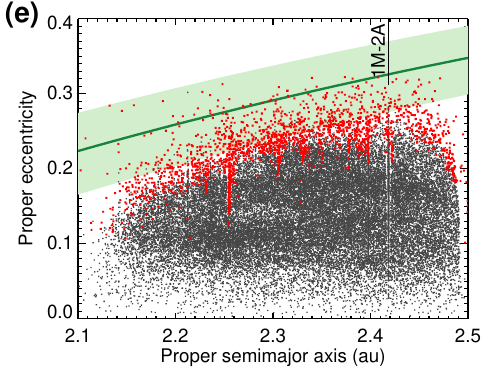}
	\includegraphics[width=1.74in]{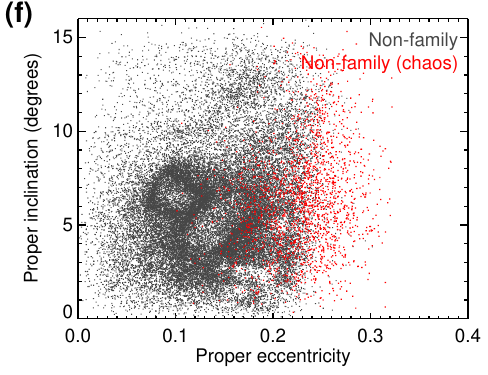}
    	\caption{Proper orbital elements of the asteroids in the IMB with $H < 16.5$ (the observational completeness limit). Panels (a)-(c) show the major asteroid families defined by \citet{NesvornyHCM3.0}. The green shaded region in panel (b) is the Mars-crossing zone, and the dashed curve in panel (a) is the $\nu_6$ secular resonance. Panels (d)-(f) show the non-family asteroids, color-coded according to the maximum Lyapunov characteristic exponent, $mLCE$ calculated by \citet{KnezevicMilani2000}, with red being the most chaotic ($mLCE > 0.00012$ per year) and black the most stable ($mLCE < 0.00012$ per year) (taken from \citealt{Dermott.et.al2021}).}
    	\label{fig:1}
\end{figure}

\begin{figure}
	\centering
    	\includegraphics[width=2.5in]{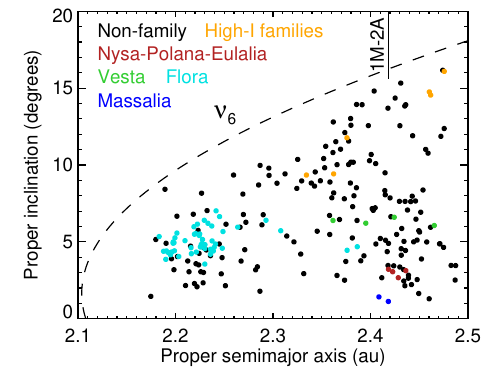}
	\caption{Scatter plot in $a-I$ space of all the asteroids in the IMB with absolute magnitude, $H < 12$. Non-family asteroids are colored black. Asteroids in the prominent Flora family are colored cyan. A CC asteroid with $H = 12$ and albedo, $A = 0.05$ has a diameter, $D=24\,km$. An NC asteroid with $H = 12$ and albedo, $A = 0.24$ has a diameter, $D=11\,km$ (taken from \citealt{Dermott.et.al2021})}
    	\label{fig:2}
\end{figure}

Fig. \ref{fig:1} shows the major families in the IMB as defined by \citet{NesvornyHCM3.0} using the HCM. For those asteroids with $H<16.5$, the IMB is dominated by family and halo asteroids and only an estimated 24\% of the asteroids are non-family and possibly primordial \citep{Dermott.et.al2022}. The largest asteroids in the major families have diameters of 525 $km$ (Vesta), 254 $km$ (Flora), 135 $km$, 142 $km$, 495 $km$ (the Nysa, Polana, Eulalia complex), 135 $km$ (Massalia), and 72 $km$ (Erigone), consistent with the argument that the planet-forming planetesimals were large with diameters $D\sim100$ $km$ \citep{Morbidelli2009}. However, the percentage of asteroids in the IMB that are non-family is size-dependent. Fig. \ref{fig:2} shows that the asteroid population with $H<12$, particularly in the range 2.3 $au$ $< a <$ 2.5 $au$, is dominated by non-family asteroids that are unlikely to be halo asteroids. For reference, a CC asteroid (defined here as an asteroid with $A<0.13$) with $H = 12$ and albedo, $A = 0.05$ has a diameter, $D=24\,km$. An NC asteroid (defined here as an asteroid with $A>0.13$) with $H = 12$ and albedo, $A = 0.24$ has a diameter, $D=11\,km$. These diameters are considerably smaller than those associated with the precursor asteroids of the major families and an outstanding question is what fraction of these non-family, non-halo asteroids are members of ghost families originating from larger primordial asteroids? 

The ages of the families have been estimated from the V-shaped spread of the asteroids in $a-1/D$ space, yielding  the result that most families have ages less than half the age of the solar system \citep{Spoto.et.al2015}. Given that catastrophic disruption was probably greater in the early solar system, this implies that a large fraction of the non-family asteroids could be remnants of ghost families. The spreading rate of the families is determined by the strength of the diurnal Yarkovsky forces and is given by
\begin{equation}
 	\frac{da}{dt}=\pm \xi \frac{3}{4\pi }\frac{1}{\sqrt{a}(1-e^2)}\frac{L_{\odot}}{c\sqrt{GM}_{\odot}}\frac{1}{D\rho }
	\label{eq:3}
\end{equation}	
where $L_{\odot}$ and $M_{\odot}$ are the solar luminosity and the solar mass, $c$ is the speed of light, and $D$ and $\rho$ are the diameter and the mean density of the asteroid \citep{Greenberg.et.al2020}. The Yarkovsky efficiency, $\xi$, depends on the spin pole obliquity and the thermal properties of the asteroid \citep{Bottke.et.al2002}. Observations of near-Earth asteroids (NEAs) show that  
\begin{equation}
 	\xi=0.12_{-0.06}^{+0.16}
	\label{eq:4}
\end{equation}
\citep{Greenberg.et.al2020}. Including the uncertainties in the mean densities, the total uncertainties in the evolution rate  may be as large as a factor of two and therefore it is possible that the family ages span the age of the solar system. It would be useful to have estimates of the strengths of the Yarkovsky forces for the separate CC and NC main belt asteroids that do not depend on estimates of the thermal parameters and mean densities, etc. Direct observations of the changes in the asteroid orbits are not yet available for main belt asteroids and we must look to other observations. It has been shown that the Martian 1:2 mean motion resonance has a well-defined excess of asteroids \citep{Gallardo2011, Dermott.et.al2022}. This excess depends on the rate of orbital evolution and modeling could yield an estimate of the strength of the Yarkovsky forces. However, that modeling has yet to be  undertaken.

\begin{figure}
	\centering
    	\includegraphics[width=1.74in]{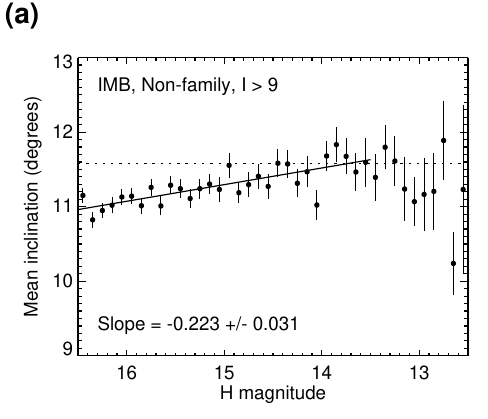}
	\includegraphics[width=1.74in]{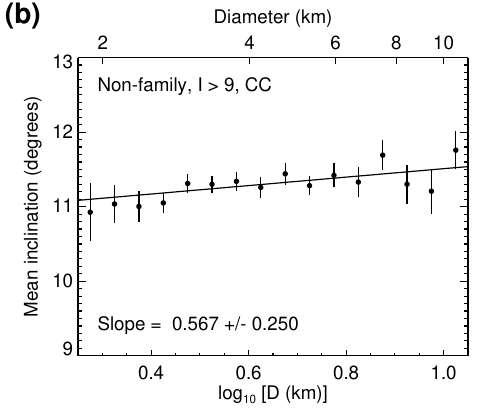}
	\includegraphics[width=1.74in]{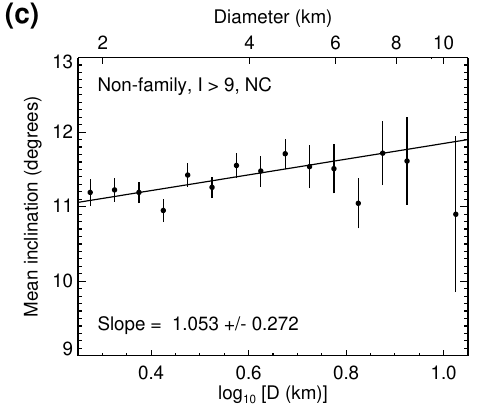}
    	\caption{Panel (a): the variation with absolute magnitude, $H$ of the mean proper inclination of the high inclination ($I>9$ $deg$), non-family asteroids. The data is shown binned in $H$, but the slope has been determined from the individual points in the range $16.5 > H > 13.5$. Panels (b) and (c): in these plots we have divided the data into CC and NC groups as determined by their WISE albedos (taken from \citealt{Dermott.et.al2021}).}
    	\label{fig:3}
\end{figure}

The outer half of the IMB (Fig. \ref{fig:2}), the half that is not dominated by the Flora family, has given us one window into the non-family asteroid population. A second significant window is given by the $a-I$ distribution shown in Fig. \ref{fig:1}a. Here, we see that the $I>9$ $deg$ region is devoid of major families and the halo asteroids associated with those families. Fig. \ref{fig:3}a shows that for the non-family asteroids in this window there is a strong (7$\sigma$) correlation between the asteroid magnitude, $H$ and the proper inclination, $I$ \citep{Dermott.et.al2021, Dermott.et.al2022}. This correlation is supported by the correlations of the separate subsets of CC (albedo, $A<0.13$) and NC (albedo, $A<0.13$) asteroids (Figs. \ref{fig:3}b, c). This correlation arises because the length of the escape route in the IMB, that is, the distance between the $\nu_6$ secular resonance and the 3:1 Jovian mean motion resonance decreases as $I$ increases with the result that small asteroids are preferentially lost from the high inclination orbits \citep{Dermott.et.al2021, Dermott.et.al2022}. To estimate the effectiveness of Yarkovsky forces in the removal of small asteroids, we can evaluate eqn. (\ref{eq:3}) and write
\begin{equation}
 	\frac{1}{a}\frac{da}{dt}=\pm 2.0\xi \left ( \frac{1\,km}{D} \right )\left ( \frac{2000\,kgm^{-3}}{\rho } \right )Gyr^{-1}.
	\label{eq:5}
\end{equation}
Given that
\begin{equation}
 	D=\frac{1329\,km}{10^{H/5}\sqrt{A}}
	\label{eq:6}
\end{equation}
and assuming that $a=2.4$ $au$, we can write
\begin{equation}
	\frac{da}{dt}=0.00043\,10^{H/5}\sqrt{A}\left ( \frac{2000\,kgm^{-3}}{\rho } \right )au\,Gyr^{-1}.
	\label{eq:7}
\end{equation}
For NC asteroids, using $A\approx 0.24$ and $\rho \approx 2000\,kgm^{-3}$ we have
\begin{equation}
	\frac{da}{dt}=0.0002\,10^{H/5}\,au\,Gyr^{-1}.
	\label{eq:8}
\end{equation}
For CC asteroids, more appropriate values would be $A\approx 0.05$ and $\rho \approx 1000\,kgm^{-3}$. However, since $\sqrt{0.24/0.05}/2\approx1$, either set of parameters results in the same estimate for $da/dt$ and we can usefully apply eqn. (\ref{eq:8}) to all of the IMB asteroids. If we assume that $da\approx0.2\,au$ typically results in the loss of an asteroid through one of the escape hatches, then asteroids with $H\gtrsim15$ will be lost from the IBM in $\lesssim1$ $Gyr$. $H=15$ corresponds to $D\approx6\,km$ (CC) or $D\approx3\,km$ (NC).

Unless the Yarkovsky efficiency, $\xi$ estimated from the NEA observations is an overestimate we must expect a significant loss of small asteroids from the IMB. This loss could be compensated for by the creation of small asteroids through the catastrophic destruction of larger asteroids. However, we have other observations that place constraints on the creation rate. One constraint is provided by the $H-I$ correlation shown in Fig. \ref{fig:3}a. \citet{Dermott.et.al2021, Dermott.et.al2022} have argued that the loss of asteroids from the IMB is dominated by Yarkovsky orbital evolution, and if the initial distribution of asteroids in $a-I$ space was uniform, and the loss of asteroids occurred over the age of the solar system, then $\xi$ must be a factor of three smaller than the NEA estimate or the correlation would be three times larger. Alternatively, if the orbital evolution rate is consistent with the NEA observations, then the asteroids must be younger than the age of the solar system by a factor of three, implying that the majority of asteroids in the IMB with high inclination ($I > 9$ $deg$) orbits are not primordial but members of families and the products of the catastrophic destruction of a small number of larger asteroids. A diffuse cluster of high-$I$ asteroids, centered on $I=13.5$ $deg$ and $e=0.18$ is evident in Fig. \ref{fig:1}c.

\begin{figure}
	\centering
    	\includegraphics[width=1.74in]{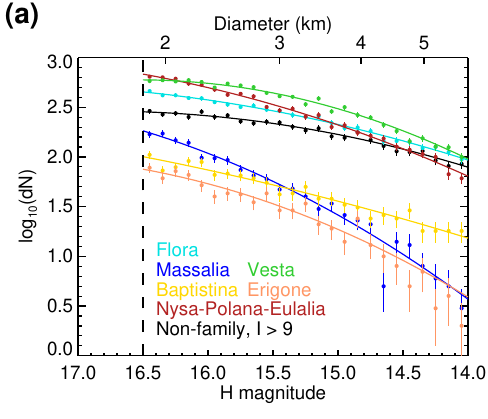}
	\includegraphics[width=1.74in]{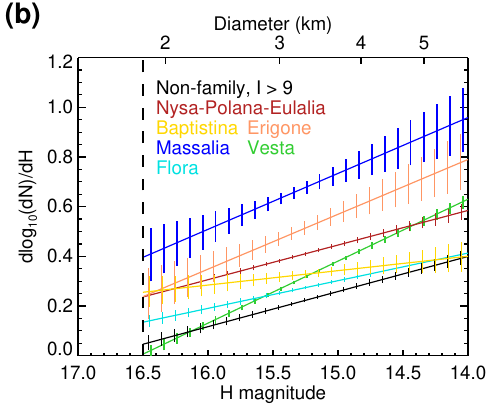}
	\includegraphics[width=1.74in]{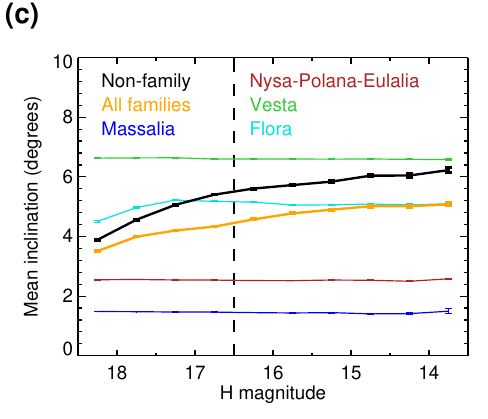}
    	\caption{Panel (a): quadratic fits to the SFDs for asteroids in the separate major families, and for the non-family ($I>9$ $deg$) asteroids. Nominal asteroid diameters have been calculated from $H$ assuming an albedo of 0.13. Panel (b): the variation with absolute magnitude, $H$ of the slopes to the quadratic fits shown in panel (a). Panel (c): the dependence of the mean proper inclination on asteroid size for all family asteroids (treated as one group), for all non-family asteroids (also treated as one group), and for the separate major families (taken from \citealt{Dermott.et.al2018, Dermott.et.al2021}).}
    	\label{fig:4}
\end{figure}

Other evidence for the net loss of small asteroids from the IMB is obtained from the size-frequency distributions (SFDs). The data in Fig. \ref{fig:4} are observationally complete and show that all the major families and the non-family asteroids in the IMB have a depletion of small asteroids. As $H$ increases the observed slopes of all the SFDs tend to zero. Modeling of the variation of the SFD with $H$ has shown that for $H\gtrsim16.5$ we must expect a marked lack of small asteroids \citep{Dermott.et.al2021, Dermott.et.al2022}. Thus, the expectation that for high $H$ the SFD tends to the equilibrium distribution postulated by \citet{Dohnanyi1969} is not supported by the observations. This is because \citet{Dohnanyi1969} considered a closed system and did not allow for the loss of small asteroids from that system. This has major consequences with respect to calculations of the expected rate of collisional disruption, and the rate of change of  asteroid spin directions due to impacts with small asteroids. Given that we do not know the SFD of the small asteroids that impact the larger asteroids, we must accept that these rates are unknown.

\section{The Mars-crossing zone}

\begin{figure}
	\centering
    	\includegraphics[width=1.74in]{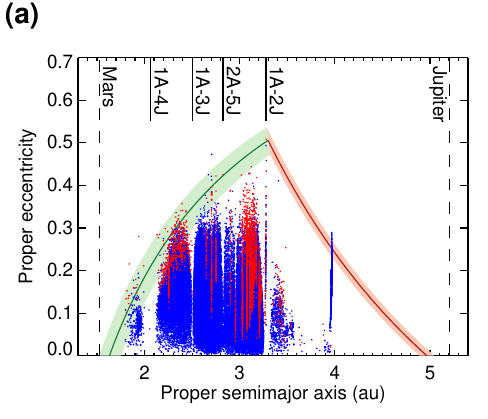}
	\includegraphics[width=1.74in]{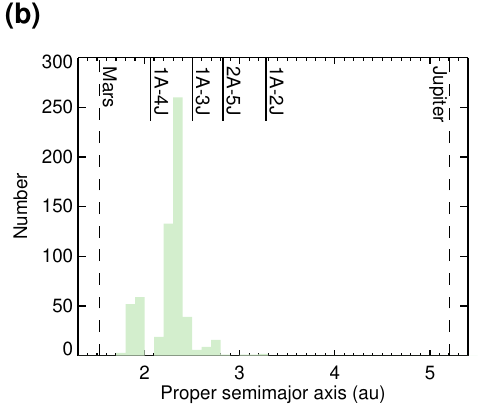}
	\includegraphics[width=1.74in]{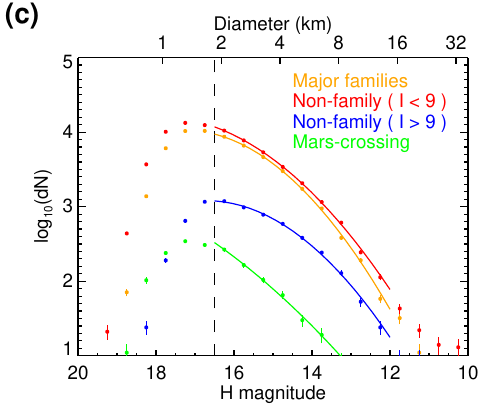}
    	\caption{Panel (a): scatter plot of the proper eccentricity $e$ and the semimajor axis, $a$ of the asteroids in the IMB with absolute magnitude $H<15$. The green shaded zone on the left is the Mars-crossing zone. Panel (b): histogram of the semimajor axes of the asteroids in the Mars-crossing zone. Panel (c): quadratic fits to the SFDs (with $dH = 0.5$) for the asteroids in the separate major families, the non- family ($I<9$ $deg$) asteroids, the non-family ($I>9$ $deg$) asteroids, and the asteroids in the Mars-crossing zone. Nominal asteroid diameters have been calculated from $H$ assuming an albedo of 0.13 (taken from \citealt{Dermott.et.al2021}).}
    	\label{fig:5}
\end{figure}

The orbital eccentricities of the main-belt asteroids are largely capped by the Mars-crossing zone (Fig. \ref{fig:5}a).   Given that the lifetime of the asteroids in this zone is $\lesssim10^8$ $yrs$, this indicates that Mars is currently scattering asteroids out of the main belt and into the inner solar system. In Fig. \ref{fig:5}b, we observe that most of the asteroids in or above the crossing-zone are in the inner main belt (IMB), suggesting that the IMB is the major source of near-Earth asteroids (NEAs) and meteorites \citep{Dermott.et.al2021, Dermott.et.al2022}. This conclusion is supported by the results of numerical investigations of the likely escape routes from the main belt \citep{Gladman1997, Granvik2017, Granvik2018}. The latter studies used estimates of the total number of NEAs after correcting for the incompleteness of the observed NEA population, estimates of the SFD of the main belt asteroids, and estimates of the strengths of the Yarkovsky forces. However, the lifetimes of the asteroids in the Mars-crossing zone are determined solely by the rate of close encounters with Mars and it is possible to determine those lifetimes without making unnecessary assumptions. The calculations for the asteroid set shown in Fig. \ref{fig:5}b have yet to be performed, but it is possible, given that corrections for observational selection and any assumptions about SFDs or Yarkovsky evolution rates are not needed, that, if the Mars-crossing population is an equilibrium population, then these calculations could give a robust estimate of the rate of transfer of asteroids from the main belt to near-Earth space.

\begin{figure}
	\centering
    	\includegraphics[width=2.5in]{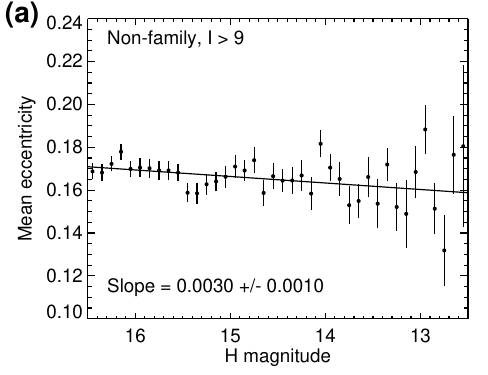}
	\includegraphics[width=2.5in]{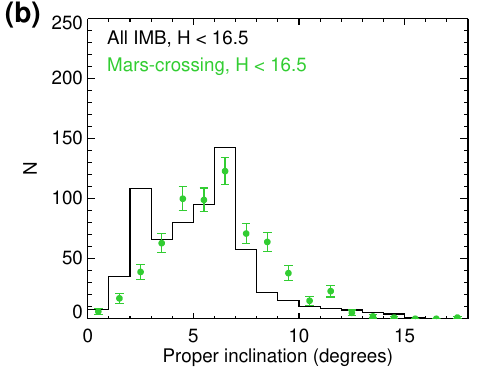}
    	\caption{Panel (a): the variation with absolute magnitude, $H$ of the mean proper eccentricity of the high inclination ($I>9$ $deg$), non-family asteroids. Panel (b): histograms of the inclination distributions of all the asteroids in the IMB with $H < 16.5$ and the asteroids in the Mars-crossing zone shown in Fig. \ref{fig:1}e. These distributions have been normalized (taken from \citealt{Dermott.et.al2021}).}
    	\label{fig:6}
\end{figure}

Asteroids are lost from the IMB through dynamical mechanisms that include chaotic and Yarkovsky driven orbital evolution. These mechanisms result in asteroids diffusing into various weak mean motion resonances (Fig. \ref{fig:1}e), and then into the Mars-crossing zone \citep{MorbidelliNesvorny1999, FarinellaVokrouhlicky1999}. A third mechanism depends on Yarkovsky forces driving small asteroids into the $\nu_6$ secular resonance \citep{Farinella.et.al1994, Migliorini.et.al1998, Farinella.et.al1998}. Numerical integrations have shown that all these mechanisms are viable. The questions that remain are which mechanism is dominant and what is the rate of loss. \citet{Greenberg.et.al2020} have shown from observations of the current orbital evolution of 247 NEAs that the ratio of retrograde to prograde spin of these objects is as high as $2.7_{-0.6}^{+0.3}$. This implies that prior to their escape these objects were evolving towards the Sun and thus that the $\nu_6$ secular resonance is the dominant escape route. There are other observations that are devoid of observational selection effects, that have yet to be analyzed, that could also place constraints on the loss mechanisms. In Fig. \ref{fig:6}a we see that the mean proper eccentricity of the high eccentricity asteroids is size-dependent, increasing with decreasing asteroid size, suggesting the action of Yarkovsky forces. Fig. \ref{fig:6}b shows the distribution of the inclinations of the asteroids in the Mars crossing zone. This distribution is somewhat similar to the distribution of the major families. There is also an excess of asteroids with high inclinations, but it has not been determined  that this excess is strong enough to support the dominant role of the $\nu_6$ secular resonance. 

\section{Vesta and the origin of the HED meteorites}

The action of Yarkovsky forces has resulted in a marked net loss of small asteroids from the IMB, implying that the remnants of many ghost families could now be sparsely distributed among other large, background asteroids. Progress with the detection of these dispersed families has been made by \citet{Delbo.et.al.2017, Delbo.et.al.2019}. Here, we show that further evidence for the loss of small asteroids from old families is provided by an analysis of  the SFD of the Vesta family.

One third of the $\sim$29,000 asteroids in the IMB with $H<16.5$ originate from Vesta. This asteroid is the source of the HED meteorites and this is the only firm asteroid-meteorite link \citep{McCord.et.al1970, McSweenBinzel2022}. Vesta is the largest asteroid in the IMB and has remained intact since its formation, thus the asteroids in this large family must originate from one or more craters. Over 25 years ago, Hubble observations revealed an enormous crater, Rheasilvia, and this crater is thought to be the source of the meteorites \citep{Thomas1997}. But later the Dawn spacecraft revealed a second enormous crater, Veneneia, partly underneath Rheasilvia \citep{Schenk2012}. Given that Veneneia has diameter of $\sim400$ $km$, its formation would probably have resulted in the formation of a large family. Is it possible that Vesta is currently associated with two families of asteroids, of different ages, that have sampled different surface locations and mantle depths?

\begin{figure}
	\centering
    	\includegraphics[width=1.76in]{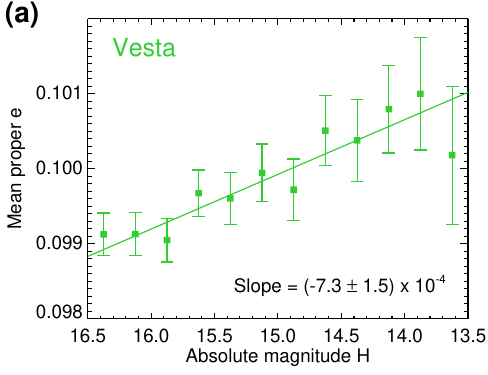}
	\includegraphics[width=1.76in]{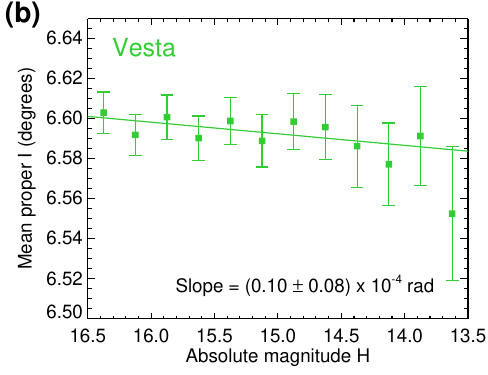}
	\includegraphics[width=1.74in]{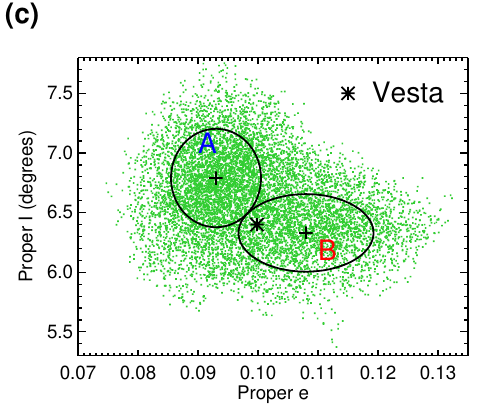}
	\includegraphics[width=1.74in]{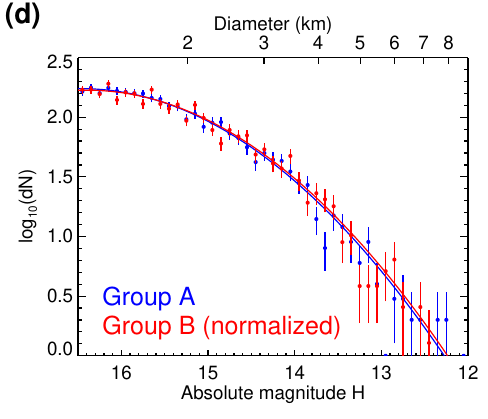}
	\includegraphics[width=1.74in]{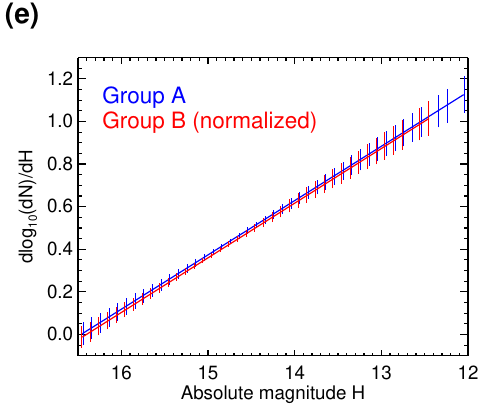}
    	\caption{Panels (a) and (b) show the variations with absolute magnitude, $H$ of the mean proper $e$ and the mean proper $I$ of the asteroids in the Vesta family. Panel (c) defines two independent regions of the Vesta family in $e-I$ space. Panel (d) shows the SFDs of the two regions shown in panel (c) after normalization. Panel (e) shows the variation with absolute magnitude, $H$ of the slopes to the quadratic fits of the SFDs shown in panel (d).}
    	\label{fig:7}
\end{figure}

We have previously argued \citep{Dermott.et.al2018} that the merging of two families with different mean proper orbital elements and different SFDs will give rise to correlations between the sizes and the mean proper orbital elements of the family members (treated as a whole) – see Fig. \ref{fig:4}c. In Fig. \ref{fig:7}a we observe a strong correlation (4.7$\sigma$) between the eccentricities and sizes of the Vesta family members. This observation, plus the asymmetric distribution of the asteroids in $e-I$ space (Fig. \ref{fig:7}c), suggests the current existence of two families. However, if we isolate two different regions of the Vesta family in $e-I$ space, as shown in Fig. \ref{fig:7}c, we observe, after normalization, that while the two regions have different mean orbital elements, the SFDs of the two regions are identical. This implies that the Vesta family must originate from one large crater and, while we believe that the Veneneia crater must have resulted in the formation of a large asteroid family, we can only account for the similarity of the SFDs if a majority of the small asteroids in the older Veneneia family have been lost from the system due to the action of Yarkovsky forces. This argument allows us to place bounds on the age of the Rheasilvia crater and on the strength of the Yarkovsky forces. Full details of these arguments will be published elsewhere.

\section{Defects of the HCM and their consequences}

What is the origin of the strong correlation observed between the mean eccentricities and sizes of the Vesta family members? Here, we suggest that this correlation may not have a physical explanation, but may be due to a fundamental flaw in the way that the HCM defines family membership. The total distance of the separation, $d$ used in eqn. (\ref{eq:1}) to define family membership gives approximately equal weights to the $\Delta a/a$, $\Delta e$ and $\Delta I$ components represented by $d_1$,  $d_2$ and $d_3$. On the initial formation of a family, these three components are determined by the structure of the debris cloud released close to the crater. It is possible that the distribution of kinetic energy between the particles in this cloud is size-dependent \citep{Leinhardt.et.al2015}. However, the second step in family formation requires these particles to reaccumulate into rubble-pile asteroids and it is not obvious that any size-energy correlation will be carried over to the resultant rubble-piles.

\begin{figure}
	\centering
    	\includegraphics[width=1.76in]{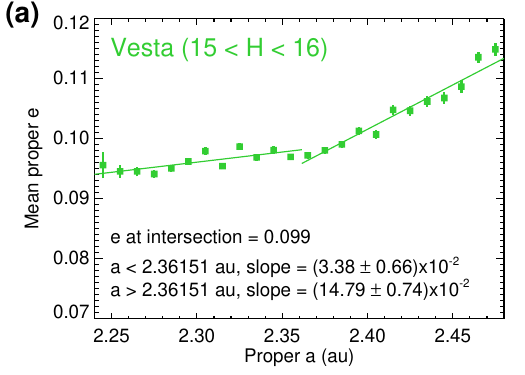}
	\includegraphics[width=1.76in]{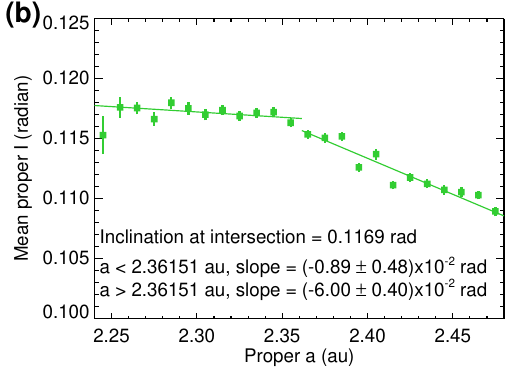}
	\includegraphics[width=1.76in]{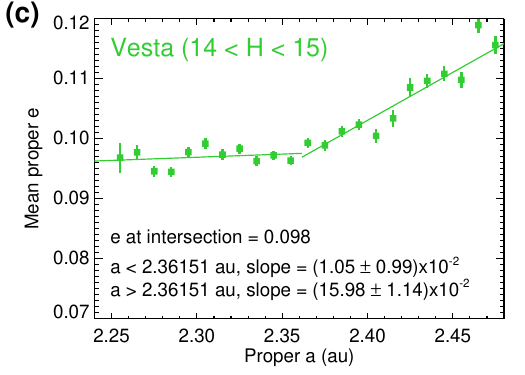}
	\includegraphics[width=1.76in]{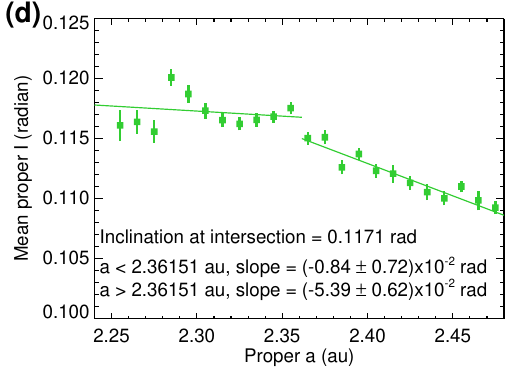}
    	\caption{Variations of the mean proper $e$ and the mean proper $I$ with semimajor axis for asteroids in the Vesta family defined by \citet{NesvornyHCM3.0} using the HCM. These variations are shown for two independent size ranges.}
    	\label{fig:8}
\end{figure}

The two components of $d$ that depend on $\Delta e$ and $\Delta I$ are comparatively stable: the changes due to Yarkovsky forces are negligible, but we must expect some small changes due to chaotic orbital evolution. If the changes in $\Delta e$ and $\Delta I$ are dominated by the changes that occur while the asteroids are trapped in weak resonances, as shown by \citet{Christou2022}, then the changes in $\Delta e$ and $\Delta I$ may be size-dependent. However, in contrast, we can be certain that the changes in the semimajor axes, due to Yarkovsky forces, are large and definitely size-dependent. Thus, the final separations of the asteroids as determined by $\Delta a/a$ are decoupled from the family formation mechanism and it is questionable that they should be used in any criterion of family membership. The inclusion of $\Delta a/a$ in the definition of $d$ in eqn. (\ref{eq:1}) could lead to the criterion that separates family from non-family asteroids being size-dependent. Evidence that this is the case is shown in Fig. \ref{fig:8}, where we see that there are strong correlations between the mean proper $e$ and mean proper $I$ and the semimajor axes of the Vesta family asteroids.

\begin{figure}
	\centering
    	\includegraphics[width=2.5in]{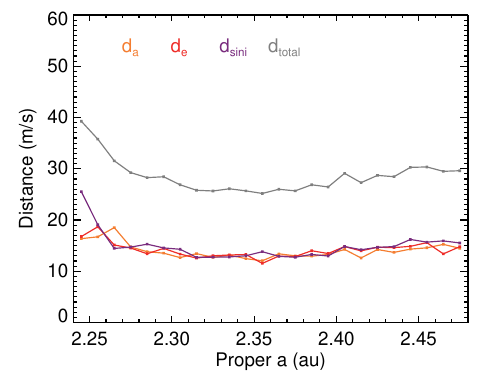}
	\caption{The variation with semimajor axis of the total and the three components of the mean minimum separation of the Vesta family asteroids in $a-e-I$ space determined using the HCM}
    	\label{fig:9}
\end{figure}

\begin{figure}
	\centering
    	\includegraphics[width=2.5in]{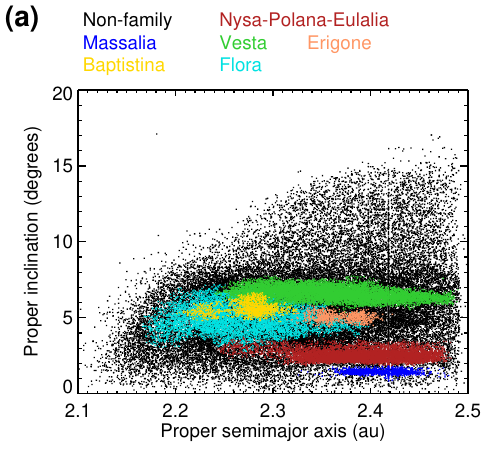}
	\includegraphics[width=2.5in]{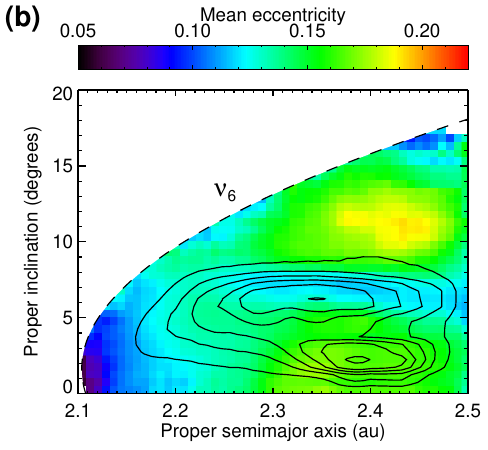}
    	\caption{Comparison of the structure of the Vesta family as revealed by two different methods. Panel (a) shows the major families as defined by \citet{NesvornyHCM3.0} using the HCM. The contours in panel (b) show the variation of number density, while the small cells show the variation of the mean proper eccentricity, $e$ in $a-I$ space.}
    	\label{fig:10}
\end{figure}

In Fig. \ref{fig:9}, we show how the mean separations of the asteroids originating from Vesta at $a=2.36$ $au$ increase with increasing distance from Vesta. We conclude from this that if the criterion for family membership, $d<d_{\textrm{crit}}$, is fixed but the mean separation of the asteroids in $a-e-I$ space increases with increasing separation from Vesta, then the number of asteroids classified as family members will decrease with increasing separation from Vesta, while the number of non-family members that are halo asteroids will increase. This is shown to be the case in Fig. \ref{fig:10}. Panel (a) of that figure shows the structure of the major families in $a-I$ space as defined by \citet{NesvornyHCM3.0} using the HCM. We observe that the width of the family, $\Delta I$ decreases as the semimajor axis increases to 2.5 $au$. Panel (b) shows the variation of the mean eccentricity of the IMB asteroids in $a-I$ space. In this plot, we have not defined the families, but we show those features that allow us to recognize the families. The contours in panel (b) show the variation of the number density, and by comparing the two panels we can see that the highest number densities are associated with the Vesta family and the Nysa-Polana-Eulalia complex. Fig. \ref{fig:10}b also shows the variation of the mean proper eccentricity in $a-I$ space.

Inspection of Fig. \ref{fig:1} shows a distinct contrast between these two major families. While the Vesta family has a high mean inclination and a low mean eccentricity, the opposite is the case for the Nysa-Polana-Eulalia complex. In Fig. \ref{fig:10}b, the band of low eccentricity asteroids with mean inclination close to the inclination of Vesta is clearly associated with the Vesta family. However, while the width of the Vesta family, $\Delta I$ shown in panel (a), as determined by \citet{NesvornyHCM3.0} using the HCM, decreases as the semimajor axis increases to 2.5 $au$, the opposite is true for the width of the band of low eccentricity asteroids (i.e., the cyan-blue band between $I\sim4.5$ $deg$ and $I\sim8$ $deg$) shown in panel (b). The width of this low eccentricity band increases markedly as the 3:1 Jovian resonance is approached, suggesting that some of the asteroids in panel (a) close to $a=2.5$ $au$ that have been classified as non-family by \citet{NesvornyHCM3.0} should be classified as halo asteroids of the Vesta family. This shows a major deficiency in the ability of the HCM to separate family asteroids from the asteroids in the associated halo, and it is this deficiency that leads to the size, orbital element correlations shown in Figs. \ref{fig:7} and \ref{fig:8}.

As an aside, the distribution of the mean proper eccentricity in $a-I$ space shown in Fig. \ref{fig:10}b allows us to recognize a yellow region of high eccentricity that could be associated with the cluster centered on $I=13.5$ $deg$ and $e=0.18$ evident in Fig. \ref{fig:1}c. Because this cluster does not extend the full extent of the distance between the $\nu_6$ secular resonance and the 3:1 Jovian mean motion resonance, this cluster may be a recent family rather than a ghost family. 

\section{Conclusions}

The SFDs of the asteroids in the IMB are inconsistent with the equilibrium distribution postulated by \citet{Dohnanyi1969}. The separation of the family and halo asteroids achieved using the HCM is somewhat inadequate and this could lead to an overestimate of the depletion of small family asteroids. However, this is not a large concern because Fig. \ref{fig:5}c shows that both family and non-family are depleted in small asteroids. If this depletion applies to the main belt as a whole, then while the SFD of the target asteroids may be well constrained, the SFD of the smaller bullets is uncertain, and therefore we must accept that the collisional lifetimes of the asteroids are uncertain, that the rate of delivery of small asteroids from the inner belt to the inner solar system is also uncertain, and that the rate of production of small asteroids and the associated NEAs and meteorites may be a stochastic process and time variable. This conclusion is consistent with previous arguments on the origin of the solar system dust bands. Immediately after the discovery of those bands by IRAS, two theories of their origin were proposed. The first theory postulated that the dust bands were an equilibrium feature associated with the three most prominent asteroid families: Eos. Themis and Koronis \citep{Dermott.et.al1984}. The second theory postulated that they were random features associated with single collisions and not necessarily related to the major families \citep{SykesGreenberg1986}. Dynamical modeling of the structure of the “ten-degree band” showed that the initial mean proper inclination of the dust particle orbits needed to account for that band is 9.3 $deg$ and equal to the mean proper inclination of the Veritas family asteroids rather than that of the Eos family \citep{Grogan2001, Dermott.et.al2002}, thus favoring the stochastic model.

\end{document}